\documentclass[prl,aps,twocolumn,groupedaddress,showpacs,floatfix,final,superscriptaddress,nofootinbib,floatfix]{revtex4-1}
\usepackage{amsmath,amssymb,multirow,bm}

\newcommand{\beq}{\begin{equation}}
\newcommand{\eeq}{\end{equation}}
\newcommand{\bea}{\begin{eqnarray}}
\newcommand{\eea}{\end{eqnarray}}
\usepackage{graphicx}
\begin{document}

\title{And there was light}


\author{C.A. Bertulani\email{carlos.bertulani@tamuc.edu}}
\affiliation{Department of Physics and Astronomy, Texas A\&M University-Commerce, Commerce TX 75429, USA}

\begin{abstract}
I discuss the use of light as a collection of real and virtual photons to study some lingering questions in particle and nuclear physics. 
\end{abstract}

\maketitle

\section{Light}

Astronomers have relied their observations almost entirely on light, although other particles have also been used as stellar tale-tellers \cite{Bah89} on a much smaller scale. Light is also a popular tool in other areas of science, arguably the most popular one. By light I mean electromagnetic radiation of all frequencies $\omega$ and wavelengths $\lambda$, such as infrared, visible light, ultraviolet, microwaves, x-rays and gamma-rays. It was only about a little more than one century ago that Planck \cite{Pla00} and  Einstein \cite{Ein05} realized that light is made of a collection of particles called photons\footnote{I cite Planck and Einstein here with the hope that their h-index and number of citations increase enough to compete with the likes such as Witten and Maldacena.} and that their wavelength and frequency are related by $\omega = 2\pi c/\lambda$, with $c$ being the speed of light. Photons are wave packets with energy $E$ and momentum $p$ content given in terms of their frequency by $E=\hbar \omega=pc$ with  $\hbar = 6.582\times 10^{16}$ eV.s being the Planck constant. Note that a 100 watt light bulb emits about $10^{15}$ photons/cm$^2$/s at 1 meter distance and a typical 100 watts FM radio antena in the frequency of $10^8$ hertz emits about $10^{12}$ photons/cm$^2$/s at about 100 kilometers. Thus, light behaves as classical waves for most practical purposes due to the large number of photons \cite{Jac98}.  

In quantum mechanics, the ``bare" wavefunction of the photon can be written as\footnote{From now on, I use the convenient unit system in which $\hbar = c =1$.}
\begin{equation}
\left| \gamma _{bare}\right\rangle = \vec{\xi}_{\bf k}\exp(i{\bf k}\cdot {\bf r}), \label{bareph}
\end{equation}
where $\vec{\xi}$ is the polarization vector associated with the photon helicity (spin = $\pm 1$), i.e. spin projection along the photon momentum ${\bf k}=\omega \hat{\bf k}$. When the photon interacts with matter, i.e. with other particles, it induces processes which can be quantified in terms of a cross section. The photon electromagnetic field at the position of the particles can be expanded into multipole contents, something similar to a McLaurin expansion of a function of position. In the case of the photon wave function above, this amounts to expand the exponential in Eq. \ref{bareph} as $1 +i{\bf k}\cdot {\bf r} - ({\bf k}\cdot {\bf r})^2/2 + \dots$, or more precisely in terms of a sum in spherical Bessel functions $j_l(kr)$. The product of this expansion with the polarization (or helicity) vector can be arranged in a similar form but with a more complex function of momentum and position, $\bf k$ and $\bf r$.  One gets the sum $ \sum_{E/M,lm} {\bf F}_{E/M,lm}({\bf k},{\bf r})$ where ${\bf F}$ involves the functions $j_l(kr)$ and depends on the electric ($E$) and  magnetic ($M$) contents of the photon field. This is known as a multiple expansion of the photon field with $l=0,1,2, \cdots$ and $m=-l, -l+1, \cdots, l-1, l$. The  cross section for the interaction of real photons with particles is given by
\begin{equation}
\sigma_{\gamma}  = \sum_{E/M,l} \sigma_{\gamma}^{(E/M,l)}, \label{sigg}
\end{equation}
where 
\begin{equation}
\sigma_{\gamma}^{(E/M,l)}={(2\pi)^3 (l+1) \over l[(2l+1)!!]^2} \sum_f  k^{2l+1} \left| \left\langle f || {\bf F}_{E/Ml} || i \right\rangle \right|^2 . \label{siggf}
\end{equation}
The sum in Eq. \ref{siggf} is over al possible final states $f$ and includes phase-space factors associated with the density of final states. The reduced matrix elements $\left\langle || \cdots || \right\rangle$ of the electric multipole fields take into account the transition of the initial $i$ to the final states $f$, include a sum over magnetic quantum numbers $m$ and an average over the initial spins of the system.

\begin{figure}
  \includegraphics[width=\columnwidth]{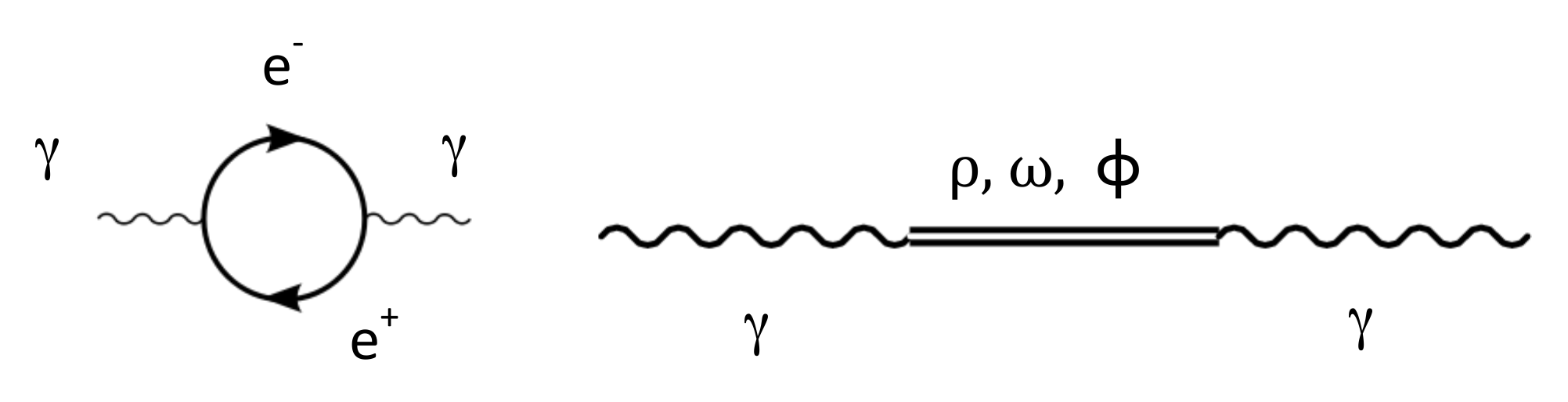}
  \caption{The photon wave function contains projections onto several particles due to quantum fluctuations.}\label{light}
\end{figure}

In the mid 1930s, Yukawa proposed a theory for nuclear forces in which the nuclear interaction is mediated by the exchange of a particle now called pion \cite{Yuk35}. This theory and the understanding that such a particle is in a virtual state was crucial for the later development of quantum field theories for electrodynamics, the strong and  weak interactions \cite{BjDr64,PeSc95,Zee00}. Yukawa's theory is the precursor of what is now known as the standard model of particle physics, which recently proved once more its incredible predictive power with the discovery of the elusive Higgs boson.  The standard model is believed to explain almost everything we know about fundamental particles.

The Coulomb potential due to a charge $Ze$ is given by $V(r)=Ze/r$ as a function of the distance $r$ from the charge. It is often more convenient to use the Fourier transform of this potential as $V(q) =(2\pi)^{-1}\int d^3 r \exp(i{\bf q}\cdot {\bf r}) V(r) = 4\pi Ze/q^2$, where $q$ is called the momentum transfer.  Following Yukawa's idea, virtual photons of momentum $q$ are thought to mediate the electromagnetic interaction between charged particles. They do not obey the Planck-Einstein energy-momentum relations because they are not observed; they are off-energy-shell. Yukawa's idea lead to the development of quantum field theories, such as quantum electrodynamics, or QED for short. In this theory, charged particles interact with others and with themselves via the emission and absorption of virtual photons: every emitted photon  is absorbed either by another particle or by itself, so that we never see them. Weird as it might look, it explains phenomena in nature, such as atomic levels, with an incredible precision, sometimes to 1 part in a trillion or better. QED is now the basis of more advanced quantum field theories (QFT) in which bosons (particles with integer spins, e.g., photons, gluons, mesons) are exchanged between  electron, quarks or between composite particles  (e.g., nucleons, hyperons).  Due to our current understanding of QFT we view the photon as a complicated object: it can fluctuate into other particles by emitting and re-absorbing them and its wave function carries imprints of such particles. That is, 
\begin{eqnarray}
\left| \gamma \right\rangle &=& C_{bare} \left| \gamma _{bare}\right\rangle +  C_{ee} \left| e^{-}e^+ \right\rangle+\cdots + C_{qq} \left| q\bar{q} \right\rangle \nonumber \\ 
&+&  C_\omega \left| \omega\right\rangle + C_\phi \left| \phi\right\rangle +C_\rho \left| \rho\right\rangle +\cdots 
\end{eqnarray}
The vector dominance model proposed by Sakurai \cite{Sak60} states that for energetic photons, the photon fluctuates  mainly into mesons. Since the photon has spin-parity $J^P = 1^{-}$,  it tends to  fluctuate into a vector meson ($\rho, \omega, \phi, J/\psi$) made of quark-antiquark pairs,  as given by the last three terms of the equation above.

\section{Virtual Light}

The 1957 Nobel Laureate, T.D. Lee, felt surprised when I mentioned a curious historical event during a seminar in 2001 at the Brookhaven National Laboratory. 1938 Nobel Laureate Enrico Fermi  (Lee's PhD advisor) published the same paper twice, just in two different languages and two different journals \cite{Fer24,Fer25}. Fermi's paper(s) contained a brilliant idea but were relatively unknown compared to his other works \cite{BeB88}. To calculate the excitation and ionization of atoms by means of energetic alpha-particles, Fermi noticed that the calculations can be much simplified if the time-dependent electromagnetic field generated by the projectiles is replaced by an equivalent pulse of real photons incident on the atom. In mid 1930's, the 1922 Nobel Laureate Niels Bohr proposed to then young physicist Carl Friedrich von Weisz\"acker a method to improve Fermi's idea by including relativistic corrections. The equations obtained by Weisz\"acker \cite{Wei34} and independently by Williams \cite{Wil34} contain the Lorentz contraction factor $\gamma = (1-v^2)^{-1/2}$ in some places of Fermi's original formulas ($v$ is the projectile velocity). For this reason, Fermi's method became widely known in textbooks and scientific publications as the Weizs\"acker-Williams method.  This is a a bit unjust. But Fermi was Fermi and this event did not make any difference in his career\footnote{Except for including one double-counting in his  publications list. But, as late physicist Hermann Feshbach used to say: ``It is better double-counting than no counting".}. 

Fermi's method is summarized in a simple equation
\begin{equation}
\sigma = \int d\omega {n(\omega) \over \omega} \sigma_\gamma (\omega), \label{epn}
\end{equation}
where $\sigma$ is the cross section for the process under consideration induced by the charged projectile, $ \sigma_\gamma$ is the cross section for the same process generated by a real photon, and $n(\omega)$ is the number of equivalent photons incident on the target. The integration runs over all photon energies in the light pulse. Fermi's virtual photon idea is popular in atomic, nuclear and particle physics, being described in classical \cite{Jac98,PeSc95} and quantum mechanics  textbooks. 

The expressions for the number of equivalent photons (EPN) derived by Fermi, Weizs\"acker and Williams have been improved along the years to include the virtual photon multipolarity \cite{BeBa85,BeNa93}, projectile spin content \cite{Ber07}, and collision geometry corrections \cite{CaJa90, BaFe90}.  Expression \ref{epn} can now be written as \cite{BeB88}
\begin{equation}
\sigma = \sum_{E/M,l}\int d\omega {n_{E/M,l}(\omega) \over \omega} \sigma_\gamma^{(E/M.l)} (\omega). \label{epn2}
\end{equation}
This expression is exact in QED in the one-photon exchange approximation and when the projectile and target charge distribution does not overlap during the collision \cite{BeB88}. It has been very useful to study the excitation of giant resonances in relativistic nucleus-nucleus collisions. The departure from the one-photon exchange limit requires a detailed description of multiple photo-nuclear interactions which has enabled the prediction of multiphonon giant resonances, e.g., the double giant dipole resonance \cite{BePo99, Aum98}. 

Since  the EPNs given by $n_{E/M,l}$ depend on the multipolarity $E/M,l$,  the cross sections for photo nuclear processes induced by the projectile are not directly proportional to the photo nuclear processes induced by a real photon, as in Eq. \ref{epn}. However, at ultra-relativistic energies one can show that  $n_{E1}\simeq n_{E2} \simeq \cdots \simeq n_{M1}\simeq\cdots$, and Fermi's formula, Eq. \ref{epn}, is recovered \cite{BeB88}.

\begin{table*}[t]
\begin{tabular}
[c]{llllll}\hline\hline
Mesons $\eta,$ $\chi$ and $h$ $(c\overline{c})$ & $J^{PC}$ & $\Gamma
_{\gamma\gamma}^{th}$ [keV] & $\Gamma_{\gamma\gamma}^{\exp}$ [keV] & Obs. &
$\sigma_{\gamma\gamma}^{X}$\\\hline
$\eta_{c}$ & (0$^{-+}$) & 3.4 - 4.8 & $6.7_{-0.8}^{+0.9}$ & $m_{c}=1.4-1.6$
GeV & 0.26 - 0.34 mb\\
$\eta_{c}(3790)$ &  & 1.85 - 8.49 & $1.3\pm0.6$ & $m_{c}=1.4$ GeV & 0.06 - 0.1
mb\\
$\eta_{c}^{\prime}(3790)$ &  & 3.7 & unknown & $m_{c}=1.4$ GeV & 0.11 mb\\
$\eta_{c}(4060)$ &  & 3.3 & unknown &  & 0.09 mb\\
$\eta_{c2}^{1D}(3840)$ &  & $20.\times10^{-3}$ & unknown &  & 0.15 $\mu$b\\
$\eta_{c2}^{2D}(4210)$ &  & $35.\times10^{-3}$ & unknown &  & 0.14 $\mu$b\\
$\eta_{c4}^{1G}(4350)$ &  & $0.92\times10^{-3}$ & unknown &  & 0.08 $\mu$b\\
$\chi_{2}$ & (2$^{++}$) & 0.56 & $0.258\pm0.019$ & $\left(  \lambda=2\right)
/\left(  \lambda=0\right)  =0.005$ & 82 $\mu$b\\
$\chi_{0}$ & (0$^{++}$) & 1.56 & $0.276\pm0.033$ & $\Gamma_{\gamma\gamma}$
$\left(  \chi_{0}\right)  /\Gamma_{\gamma\gamma}$ $\left(  \chi_{2}\right)
=2.79$ & 0.05 mb\\
$\chi_{2}^{\prime}$ & (2$^{++}$) & 0.64 & unknown &  & 0.09 mb\\
h$_{c2}(3840)$ &  & $20\times10^{-3}$ & unknown & $^{1}$D$_{2}$ & 82 $\mu$b\\
$\chi_{2}\left(  4100\right)  $ &  & $30\times10^{-3}$ & unknown & $^{3}%
$F$_{2}$ & 0.11 $\mu$b\\\hline\hline
\label{t2} &  &  &  &  &
\end{tabular}
\caption{$\gamma-\gamma$ widths for $c\bar{c}$ mesons $\eta$, $\chi$ and $h$
calculated with  several theoretical models (for more details, see Ref. \cite{Ber09}). Experimental values of the $\gamma-\gamma$
widths are extracted form the Particle Data Properties Web site.}
\end{table*}

Fermi's method is easily generalized to the production of a particle $X$ in ultraperipheral collisions (UPC) of two charged particles.  One can write the cross section for photon-photon fusion with squared center of mass energy $s$ by \cite{BeB88},
\begin{equation}
\sigma_{X}=\int {dx_{1}\over x_1}{dx_{2}\over x_2} n_{\gamma}\left(  x_{1}\right)  n_{\gamma}\left(
x_{2}\right)  \sigma^{X}_{\gamma\gamma}\left(  x_{1}x_{2}s\right)  \;,
\label{eq:two-photon}%
\end{equation} 
where $n_{\gamma}(x)$ is the distribution function (EPNs)
to find a quantum $\gamma$ with energy fraction $x$ and $\sigma
^{X}_{\gamma\gamma}\left(  x_{1}x_{2}s\right)  $ is the photon-photon fusion cross
section. From this expression one could determine the reaction rate to produce anti-hydrogen atoms at the  Low Energy Antiproton Ring (LEAR) at CERN in 1996 \cite{Bau96} . The colliding ions produce $e^+e^-$ pairs with the positrons being captured in orbits around the antiprotons \cite{BeB88}. It was the first time that antimatter as we expect (i.e., anti-atoms) was produced in the laboratory, and the exciting news made its way to the New York Times and other world newspapers \cite{Bro96}\footnote{At the time, Gerhard Baur told me: ``I think that I am at the height of my career. It is easier to get a publication in Science or Nature than to have the results of our work reported in the NY Times or the BBC".}. The validity of the equivalent photon approximation was proven by a later measurement at FERMILAB \cite{Bla97} and comparison with a theoretical calculation \cite{BeBa98}.
Nowadays, one is intensively studying the properties of anti-atoms in ion traps \cite{Eug10,Gro10}. Production of muonic, pionic, and other exotic atoms by the coherent photon exchange between nuclei at the Large Hadron Collider (LHC) at CERN has also been investigated theoretically in Ref. \cite{BeEl10}.

\section{Virtual Light at CERN}

\subsection{Probing meson decay widths}

In quantum chromodynamics (QCD), meson spectroscopy is still looking for multiquark states
such as molecules $(q\overline{q})(q\overline{q})$, hybrid mesons
$(q\overline{q}g)$ and glueballs $(gg)$ \cite{Yao06}. The photo production of mesons is proportional to their decay widths, i.e., $\sigma_X \propto \Gamma_{\gamma\gamma}(X)$. 
Therefore, two-photon exchange in relativistic heavy ion collisions 
can contribute to the search for non-$q\overline{q}$
resonances by identifying  anomalous $\gamma\gamma$ couplings and  their energy spectrum. For example, $q\overline{q}$ meson decay widths explained within a flavor-SU(3) multiplet of $q\overline{q}$ states, yields
\begin{equation}
\Gamma_{\gamma\gamma}(f) \ :\ \Gamma_{\gamma\gamma}(a) \ :\ \Gamma
_{\gamma\gamma}(f^{\prime}) = 25 \ : \ 9 \ : \ 2.
\end{equation}
Except for minor relativistic corrections, it reproduces the experimental
data quite well.  The $\Gamma_{\gamma\gamma}$ partial widths of
resonances is useful in the identification of $q\overline{q}$
exotic states, but the absolute scale of these widths are rather sensitive to the assumptions made in calculation.
Thus, there is a true motivation to observe if predictions of ``abnormal" states
within the quark model is verified experimentally. Photon-photon (or \textquotedblleft
two-photon\textquotedblright) processes have long been studied at $e^{+}e^{-}$
colliders with this purpose \cite{Bud74}.

\begin{table*}[t]
\begin{tabular}
[c]{llllll}\hline\hline
Mesons $\eta,$ $\chi$ and $h$ ($b\overline{b}$) & $J^{PC}$ & $\Gamma
_{\gamma\gamma}^{th}$ [keV] & $\Gamma_{\gamma\gamma}^{\exp}$ [keV] & Obs. &
$\sigma_{\gamma\gamma}^{X}$\\\hline
$\eta_{b}^{1S}(9400)$ &  & $0.17\times10^{-3}$ & unknown &  & 19 nb\\
$\eta_{b}^{2S}(9400)$ &  & $0.13\times10^{-3}$ & unknown &  & 16 nb\\
$\eta_{b}^{3S}(9480)$ &  & $0.11\times10^{-3}$ & unknown &  & 14 nb\\
$\eta_{b2}^{1D}(10150)$ &  & $33.\times10^{-6}$ & unknown &  & 0.4 nb\\
$\eta_{b2}^{2D}(10450)$ &  & $69.\times10^{-6}$ & unknown &  & 0.8 nb\\
$\eta_{b4}^{1G}(10150)$ &  & $59.\times10^{-6}$ & unknown &  & 0.7 nb\\
$\eta_{b}(9366)$ & (0$^{-+}$) & 0.17 & unknown &  & 0.12 $\mu$b\\
$\eta_{b}^{\prime}$ &  & 0.13 & unknown &  & 0.17 $\mu$b\\
$\eta_{b}^{\prime\prime}$ &  & 0.11 & unknown & $s\overline{s},$ $m_{s}=0.55$
GeV fixed & 0.15 $\mu$b\\
$\chi_{b2}(9913)$ & (2$^{++}$) & $3.7\times10^{-3}$ & unknown &  & 0.09 $\mu
$b\\
$\chi_{b0}(9860)$ & (0$^{++}$) & $13.\times10^{-3}$ & unknown &  & 0.08 $\mu
$b\\\hline\hline
\label{t3} &  &  &  &  &
\end{tabular}
\caption{$\gamma-\gamma$ widths for $b\bar{b}$-mesons $\eta$, $\chi$ 
calculated with  several theoretical models (for more details, see Ref. \cite{Ber09}). Experimental values of the $\gamma-\gamma$
widths are extracted form the Particle Data Properties Web site.}
\end{table*}

At the LHC photon-photon collisions occur at
center of mass energies an order of magnitude higher than were available at
previously $e^+e^-$ accelerators, and photon-heavy ion collisions reach 30
times the energies available at fixed target accelerators. The Lorentz gamma
factor $\gamma=(1-v^{2}/c^{4})^{-1/2}$ in the collider frame is about 7000 for
p-p, 3000 for Pb-Pb collisions. Due to the ions large charges (e.g., $Z=82$)
and their short-interaction time ($\Delta t \simeq20 \gamma$ MeV), the
generated electromagnetic fields  are much
stronger ($\propto Z^{2}$) than the Schwinger critical field
\cite{Sch49,Bau08} $E_{Sch} = m^{2}/he = 1.3 \times10^{16}$ V cm. Light
particles (e.g., $e^{+}e^{-}$-pairs) are produced copiously by such fields
\cite{BeB88}.  Electroweak processes such as $\gamma
\gamma\rightarrow W^{+}W^{-}$ or the production of the
Higgs boson with such a mechanism is not negligible  \cite{Bau88,Pap89,Gra89,BeB94}.
The physics of UPCs has become object of
intensive studies in recent years
\cite{KGS97,BHT02,BKN05,BHT07,Ba08,Ba08b,Nyst08}.

Estimates of the  two-photon cross
section from Eq. \ref{eq:two-photon} can be obtained by using Low's expression \cite{Low60}, based on the detailed balance theorem,
\begin{equation}
\sigma^{X}_{\gamma\gamma}\left(  x_{1}x_{2}s\right)  =8\pi^{2}(2J+1){\frac
{\Gamma_{m_{X}\rightarrow\gamma\gamma}}{m_{X}}}\ \delta\left(  x_{1}x_{2}s
-m^{2}_{X}\right)  \label{Low}%
\end{equation}
where $J$, $m_{X}$, and $\Gamma_{m_{X}\rightarrow\gamma\gamma}$ are the spin,
mass and the two-photon partial decay width. The delta-function ensures
energy conservation.
For example, in Ref. \cite{Ber09}  the $\Gamma_{\gamma\gamma}$
widths either taken from experiment or from theory were used to generate Tables \ref{t2} and \ref{t3}.
Many values are predictive and have not been considered before. The properties of some $q\overline{q}$ states are given, and their
production cross sections are predicted for Pb-Pb collisions at the LHC.  Ion luminosities of $10^{26}$ cm$^{-2}$ s$^{-1}$ for Pb-Pb
collisions at LHC lead up to million of events (e.g. charmonium states) per
second for the largest cross sections \cite{BHT02}. The two-photon width is a
probe of the charge of its constituents, so the magnitude of the two-photon
coupling can serve to distinguish quark dominated resonances from
glue-dominated resonances (``glueballs''). The absence of meson production via
$\gamma\gamma$ fusion is a signal of great interest for glueball search. In
ion-ion collisions, a glueball can only be produced via the annihilation of a
$q\overline{q}$ pair into gluons pairs, whereas a normal $q\overline{q}$ meson
can be produced directly \cite{Ba08b}\footnote{A friend used to tell me: "When you least expect, that is when nothing happens". We know that this is not necessarily true. See,  e.g.,  Ref. \cite{BeB94}.}.

\subsection{Production of vector mesons}

Let us consider the inclusive photoproduction of 
heavy quarks ($c\bar{c}$ and $b\bar{b}$) and the 
exclusive elastic production of vector mesons ($J/\psi$ and 
$\Upsilon(1s)$) in ultraperipheral PbPb and pPb  collisions  
at $\sqrt{s_{_{NN}}}=2.76$ TeV and $5$ TeV respectively, as described in Refs. \cite{Adeluyi:2011rt,Adeluyi:2012ph,ABM12}. We investigate the sensitivity of 
photoproduction of heavy quarks  to several gluon 
distribution modifications in the medium \cite{Adeluyi:2011rt,Adeluyi:2012ph,ABM12}. 
This idea, originally proposed in Ref. \cite{Goncalves:2001vs}, can be
used to constrain parton distribution functions from data on
photoproduction of heavy quarks and of vector mesons.

Nuclear parton distribution functions, $F_{a}^A({\bf r},x,Q^2)$, are often 
expressed as a convolution of "nuclear modifications" $R_{a}^A({\bf r},x,Q^2)$ and free nucleon
parton distribution functions $f_{a}(x,Q^2)$ where the subscript $a$ denotes a
parton species and the superscript  $A$ a particular nucleus. The
variables are the position vector ${\bf r}$, parton momentum fraction 
$x$ (Bjorken-$x$), and a factorization scale $Q^2$.
The effects of nuclear modifications in 
$R_{a}^A(x,Q^2)$ can be categorized based on different intervals in $x$.
At small values of $x$ ($x < 0.04$), we have the phenomenon
referred to as shadowing, where the nuclear parton distributions are smaller compared
to the corresponding distributions in free nucleons, i.e. $R_a^A < 1$. 
Antishadowing, is an
enhancement ($R_a^A > 1$), occuring in the range $0.04 < x <
0.3$. Another depletion, the EMC effect \cite{Aubert:1983xm}, is
present in the interval $0.3 < x < 0.8$. Finally, we have another enhancement for 
$x > 0.8$, the Fermi motion region. 
It is important to note that although both shadowing and
the EMC effect (antishadowing and Fermi motion) correspond to
depletion (enhancement), the physical principles and mechanisms 
governing these phenomena are quite different. Further details can be
found in \cite{Geesaman:1995yd,Piller:1999wx,Armesto:2006ph,Kolhinen:2005az} 

The determination of parton
distributions is usually done by global fits to  
experimental data on Deeply Inelastic Scattering (DIS)
and Drell-Yan (DY) processes. 
Since gluons are electrically neutral, their distributions
cannot be directly extracted from DIS; they are inferred from sum
rules and the $Q^2$ evolution of sea quarks distributions. 
The available data is much less accessible for nuclei than for nucleons, and there is the added
complication of nuclear mass dependence. It is therefore not unusual for
nuclear gluon distributions from different fits to differ 
significantly, especially in the magnitude of the various nuclear
effects (shadowing, antishadowing, etc). This is especially obvious
at low $Q^2$.  
    
We have used four recent gluon distributions  in this study \cite{Adeluyi:2011rt,Adeluyi:2012ph,ABM12}. For
the nucleon gluon distributions we use the Martin-Stirling-Thorne-Watts
(MSTW08) parton distributions \cite{Martin:2009iq}.
In the nuclear case we use three nuclear modification sets.
Two sets are by  Eskola, Paukunnen, and Salgado, namely EPS08 and EPS09 
\cite{Eskola:2008ca,Eskola:2009uj}. 
The third is the Hirai-Kumano-Nagai (HKN07) distributions 
\cite{Hirai:2007sx}. The gluon distributions from MSTW08 serve two purposes: 
(a) as free nucleon distributions used in conjunction with
nuclear modifications, and (b) as a ``special" nuclear gluon
distribution in the absence of nuclear effects. The latter usage is
particularly useful for highlighting the influence of the various
nuclear effects on observables.
We have also used the GRV \cite{Gluck:1991jc}, SaS1D \cite{Schuler:1995fk}, and CJK2 \cite{Cornet:2004ng} resolved photon distributions. The 
characteristics of these distributions, especially the disparities in
the strength of the nuclear modifications of their gluon content, has
been treated in detail in \cite{Adeluyi:2012ph}. Our calculations are to leading order
(LO).

In UPCs with  pPb and PbPb collisions two rather different 
production mechanisms (direct and resolved) are present. In the direct  
mechanism the incident photon interacts directly with the target
(nucleus or proton) whereas in the resolved mechanism the incident
photon first fluctuates into a quark-antiquark pair (see Figure \ref{light}) which then subsequently interacts 
hadronically with the target. At leading order the direct production involves
only the gluon distributions in the target while the resolved production
requires the distributions of light quarks and 
gluons in both photon and target. The total
production cross sections and rapidity distributions are 
of course the sum of the contributions from both processes. 

\begin{figure}[!ht]
\includegraphics[width=\columnwidth]{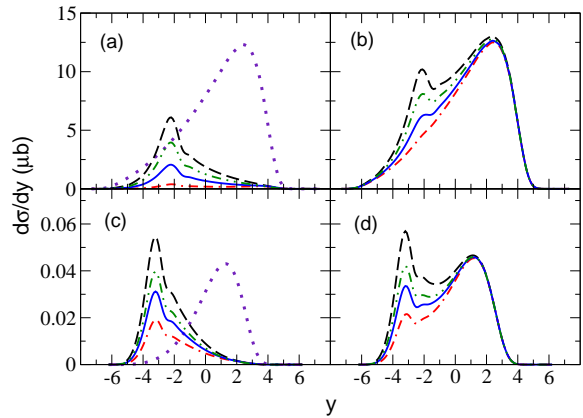}
\caption{ Rapidity distributions of 
exclusive photoproduction of $J/\psi$ (top) and  $\Upsilon$ (bottom)
in pPb collisions at $\sqrt{s_{_{NN}}}=5$ TeV.  
The left hand panels (a and c) show the $\gamma$p and $\gamma$Pb
contributions separately while the right hand panels (b and d) show the sum. 
 Dotted line depicts 
the $\gamma$p contribution while the dashed (MSTW08), dash-double-
dotted (HKN07), solid 
(EPS09), and dash-dotted (EPS08) lines correspond to $\gamma$Pb 
contributions with no shadowing, weak, moderate, 
and strong shadowing respectively.}\label{jpsiupsipPb}
\end{figure}

Let us now consider the elastic photoproduction of the $J/\psi$ 
and $\Upsilon(1s)$. As discussed in \cite{Goncalves:2001vs,Adeluyi:2011rt,Adeluyi:2012ph} 
the production mechanism for these vector mesons involves
the square of the nuclear/nucleon gluon distribution. This quadratic
dependence leads to a dramatic increase in the sensitivity of both 
cross sections and rapidity distributions to nuclear effects
(predominantly shadowing) on gluon distributions.  
In Table~\ref{tjpsiupsipPb} we present the component 
and total cross sections for the elastic photoproduction of $J/\psi$ 
and $\Upsilon$ in ultraperipheral pPb collisions at the LHC. The 
associated rapidity distributions are shown in Fig.~\ref{jpsiupsipPb}.

\begin{table}[!ht]
\caption{\label{tjpsiupsipPb} Cross sections for elastic photoproduction of 
$J/\Psi$ (in $\mu$b) and $\Upsilon$ (in nb) in ultraperipheral 
pPb collisions.}
\begin{tabular}[c]{|l|c|ccc c c c c c c c| c c c c}
\hline
&            PDF &             && $\gamma$p    &&& $\gamma$Pb    &&& Total & \\
\hline
&            MSTW08&           && 63.6        &&& 18.3         &&& 81.9 & \\
$J/\Psi$&    EPS08&            &&              &&& 1.8          &&& 65.4 & \\		       
($\mu$b)&    EPS09&	         &&              &&& 6.6        &&& 70.2  & \\		      
&            HKN07&            &&              &&& 12.0         &&& 75.6  & \\	
\hline
&            MSTW08&           && 149.6        &&& 137.0         &&& 286.6 & \\
$\Upsilon$&  EPS08&            &&              &&& 54.8          &&& 204.4 & \\		       
(nb)&        EPS09&	       &&              &&& 82.9        &&& 232.5  & \\		      
&            HKN07&            &&              &&& 101.5         &&& 251.1  & \\	
\hline                              
\end{tabular}
\end{table}
    
\begin{figure}[!ht]
\includegraphics[width=\columnwidth]{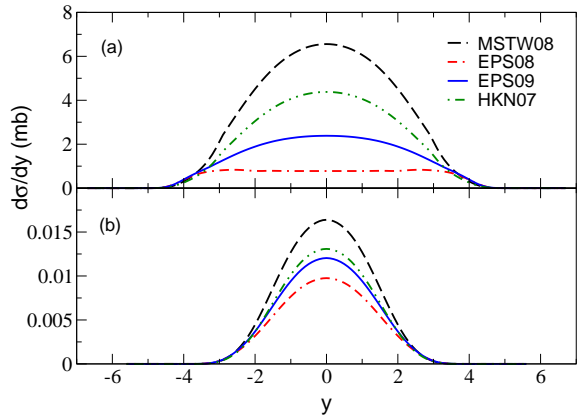}
\caption{ Rapidity distributions of 
exclusive photoproduction of (a) $J/\psi$ (top) and (b) $\Upsilon(1s)$ (bottom) in PbPb collisions at $\sqrt{s_{_{NN}}}=5.0$ TeV. 
Dashed, solid, dash-dotted, and dash-double-dotted lines are results
from MSTW08, EPS09, EPS08, and HKN07 parton distributions respectively.}\label{jpsiupsiPbPb}
\end{figure}

Unlike photoproduction of $c\bar{c}$ and $b\bar{b}$
in pPb collisions which is practically insensitive to
shadowing, the enhanced sensitivity to gluon shadowing due to the
quadratic dependence is already apparent here. Thus even though the $\gamma$p
contribution is dominant in the case of $J/\psi$ production, the
$\gamma$Pb contributions from both MSTW08 (no shadowing) and HKN07
(weak shadowing) are still relatively 
appreciable. Gluon 
shadowing present in EPS08 is enough to render its $\gamma$Pb
contribution almost negligible. For $\Upsilon(1s)$ production the 
$\gamma$Pb component contributes significantly, and is in fact
comparable to the $\gamma$p contribution in the case of MSTW08.
Due to this, the effect of gluon shadowing is more clearly reflected
in the total rapidity distributions and thus 
$\Upsilon(1s)$ production may potentially be of some use in
constraining gluon shadowing, especially in the $-4 < y < -1$ rapidity
interval. 

The results on $J/\psi$ and $\Upsilon(1s)$ production in
ultraperipheral PbPb collisions are presented in
Table~\ref{tjpsiupsiPbPb} and in
Fig.~\ref{jpsiupsiPbPb}. The differences in 
the predicted cross sections and rapidity distributions are clearly visible, especially for $J/\psi$. Thus the photoproduction of $J/\psi$
and $\Upsilon(1s)$ in ultraperipheral PbPb collisions 
are an excellent probe of gluon shadowing 
and a good discriminator of the different gluon shadowing sets.

\begin{table}[!ht]
\caption{\label{tjpsiupsiPbPb} Total cross sections for elastic photoproduction of 
$J/\psi$ (in mb) and $\Upsilon(1s)$ (in $\mu$b) in ultraperipheral 
PbPb collisions.}
\begin{tabular}[c]{|l|c c c|}
\hline
PDF                    && $J/\Psi$ (mb) & $\Upsilon$ ($\mu$b)  \\   
\hline
MSTW08              && 32.6         & 51.3  \\		      
EPS08               && 6.3          & 32.2  \\		       
EPS09	            && 13.9         & 39.0  \\		      
HKN07               && 22.1         & 41.3  \\
\hline	               
\end{tabular}
\end{table}

Let us briefly compare the present results to those
at higher collision energies (pPb at  $\sqrt{s_{_{NN}}}=8.8$ TeV and 
PbPb at $\sqrt{s_{_{NN}}}=5.5$ TeV) presented in \cite{Adeluyi:2012ph}.
For pPb collisions the cross sections for $c\bar{c}$ ($b\bar{b}$) production at 
$\sqrt{s_{_{NN}}}=8.8$ TeV are approximately $1.8$ ($2.1$) times those
at $\sqrt{s_{_{NN}}}=5.0$ TeV. The relative $\gamma$Pb
contributions are almost equal and nuclear effects are practically the 
same at both energies for both heavy quarks. 
For PbPb the cross sections at $\sqrt{s_{_{NN}}}=5.5$
TeV are approximately $2.1$ ($2.8$) times those at $\sqrt{s_{_{NN}}}=2.76$ TeV
for $c\bar{c}$ ($b\bar{b}$). Nuclear effects are about $27\%$ larger
for $c\bar{c}$ although shadowing trends are identical. 
The case of $b\bar{b}$ is interesting: strong influence of
antishadowing results in both  EPS08 and EPS09 $b\bar{b}$ cross
sections at $\sqrt{s_{_{NN}}}=2.76$ TeV being larger than that of MSTW08. 
This contrast with the behavior at $\sqrt{s_{_{NN}}}=5.5$ TeV where the
usual shadowing trend prevails. 

\begin{figure}[!ht]
\includegraphics[width=\columnwidth]{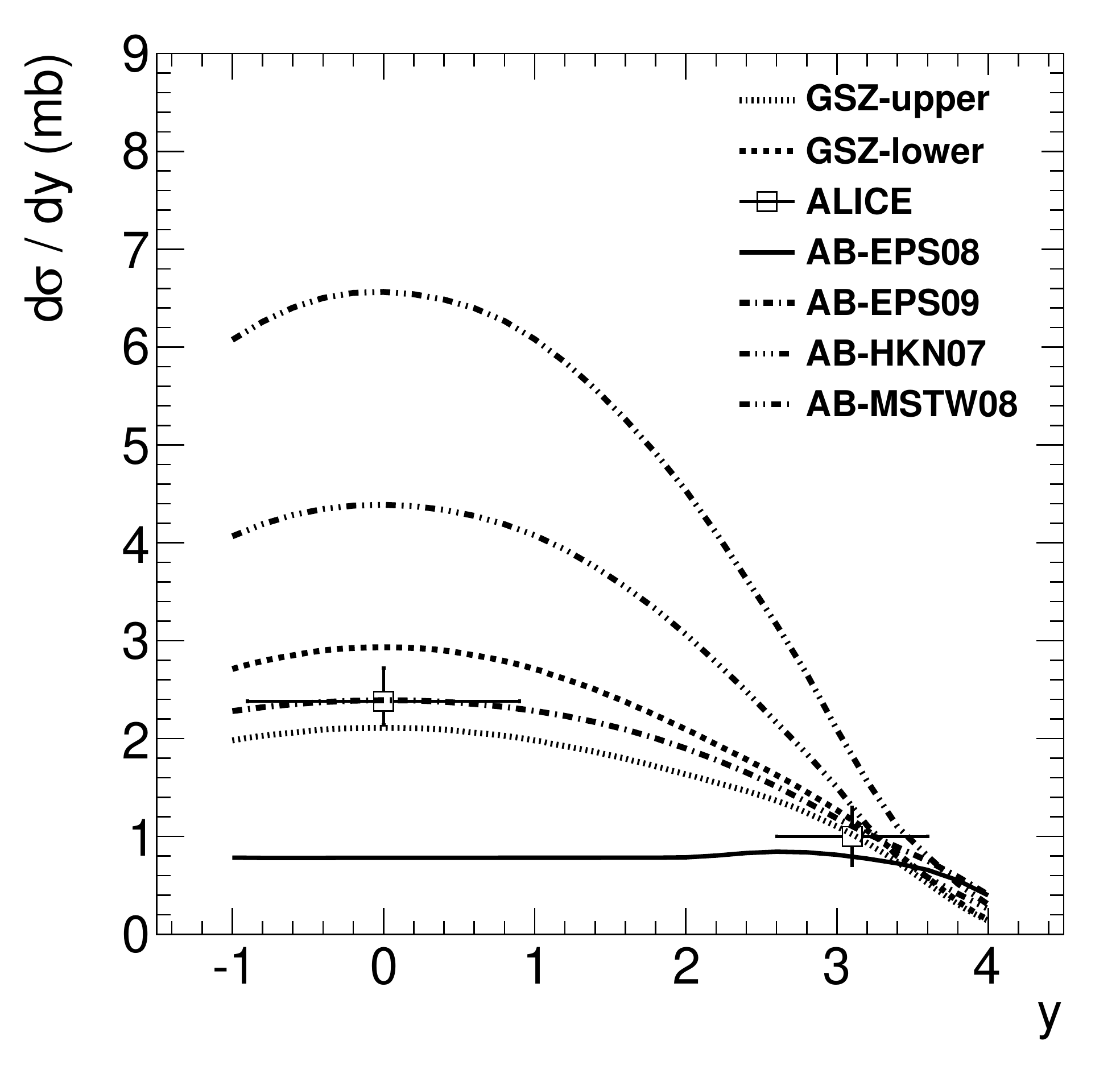}
\caption{Comparison among the published value of the cross section at forward rapidity, the result at central rapidities and several theoretical models. 
The error is the quadratic sum of the statistical and systematic errors. (Courtesy: Magdalena Malek)}\label{jpsiexpPbPb}
\end{figure}

Further differences manifest in photoproduction of $J/\psi$ and 
$\Upsilon(1s)$ in pPb collisions. For  $J/\psi$ the relative 
$\gamma$Pb contributions at $\sqrt{s_{_{NN}}}=5.0$ TeV are about twice those at
$\sqrt{s_{_{NN}}}=8.8$ TeV  and about $30\%$ larger for
$\Upsilon(1s)$. Shadowing effects are therefore more pronounced at 
lower energy and consequently better suited for constraining purposes.
As expected total cross sections for $J/\psi$ ($\Upsilon(1s)$) 
are approximately a factor of $2.5$ ($2.2$) 
larger than at $\sqrt{s_{_{NN}}}=8.8$ TeV. For PbPb collisions although
the cross sections are larger at $\sqrt{s_{_{NN}}}=5.5$ TeV, shadowing 
effects are almost the same (for $J/\psi$) or slightly larger 
(for $\Upsilon(1s)$) than at $\sqrt{s_{_{NN}}}=2.76$ TeV. Thus the  
constraining possibilities at both energies are almost equal.

The experimental results of Refs. \cite{Abe13,Abe13b,Gru14} for $J/\Psi$ production at forward and central rapidities are shown and compared to our predictions in Fig. \ref{jpsiexpPbPb}. The $J/\Psi$ production cross section is in a good agreement with  gluon distributions which incorporate the nuclear gluon shadowing  (In the figure AB stands for Refs. \cite{Adeluyi:2011rt,Adeluyi:2012ph} and AB-EPS09 refers to the EPS09 gluon distribution). Also shown are predictions from Ref. \cite{GSZ14}, denoted by GSZ. It appears that models which do not include  nuclear gluon shadowing are inconsistent with the ALICE experiment. Models which use rescattering  effects such as the AB-HKN07and AB-EPS08 models contain either too little or too much shadowing. Based on these results, perhaps confirmed by the ongoing analysis of pPb experiments, the $J/\Psi$ photoproduction cross sections at CERN can become a powerful tool to constrain nuclear gluon shadowing in the Bjorken-$x$ region of $x<10^{-3}$. Experiments with $\Upsilon$ production are also of great interest and presently under scrutiny.

\section{Light on CERN's future}

It is rewarding to realize that much of what was predicted for the physics of photo-nuclear and photon-photon processes using relativistic heavy ion collisions \cite{BeB88,Bau88,BeB94} such as pair production, with and without capture, light-by-light scattering or Delbr\"uck scattering, exotic meson production, beam depletion, and several other process are now being pursued at CERN.  It was an arduous way to convince the community to pursue such efforts since the first attempts to popularize this branch of physics \cite{BeB88,Bau88,BeB94}. 

The LHC involves a collaboration of about 10,000 scientists from more than 100 countries and  hundreds of universities and laboratories. Its  tunnel has a circumference of 27 kilometers and rests 175 meters beneath the ground. Arguably, the construction of the LHC and the discovery of the Higgs boson cost about 13 billions dollars. The annual total budget for the experiments rans over 1 billion dollars a year. Such amount of money is only justifiable when great questions of science wait for answers.  Evidently, CERN was not built with aim to study UPCs. But since the LHC is there and might be idle during some time, why not? Moreover, some of the physics described here has made their way to the news and attracted interest of a large number of physicists. So, why not do it? 

\begin{figure}[!ht]
\includegraphics[width=\columnwidth]{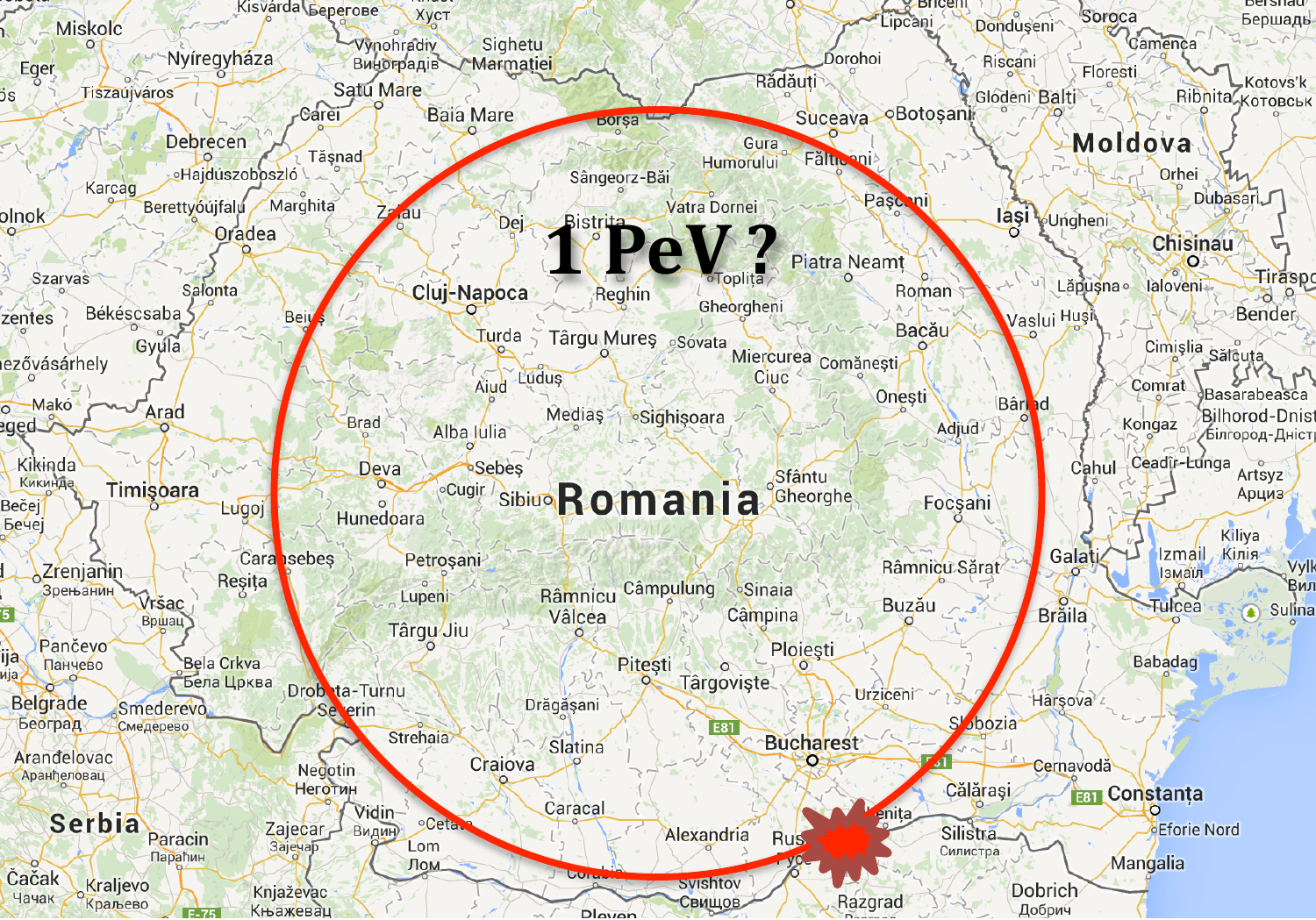}
\caption{PeV accelerator as large as the country of Romania. Future of particle physics? (Courtesy: Livius Trache)}\label{pev}
\end{figure}

During my passage for many physics departments around the world, I have heard that nuclear physics is not a fundamental science, but that particle physics is. What a particle physicist means by fundamental are answers for interactions, particles and fields, such as the Higgs, which fill the gaps of theories for forces and matter in the Universe. For those critics, nuclear physics is engineering with nucleons. Evidently, questions such as what is the origin of elements \cite{Bur57}, or the prediction of the Hoyle state in $^{12}$C and its consequence for the existence of life \cite{Hoy54,Meis13}, have also to be considered as fundamental. It looks to me that if supersymmetric particles are not found in CERN during the next five years, particle physicists will have to content themselves with precision calculations and measurements, or looking for less ``fundamental" physics such as exotic mesons, or medium modifications of parton distributions and the laboratory might have to diversify its science reach. 

There was a compelling argument to build the LHC and maybe its investment has been well justified. But somehow in the mind of some of our colleagues, there is the need to  build an even larger accelerator. I hear about a certain PeV accelerator needed for ``new physics" which might have a circular diameter as big as the country of Romania, as shown in Figure  \ref{pev}. I have attended talks on such projects and they often come accompanied with some political statements such as "CERN has wonderful teacher education programs". Needless to say that such programs are relevant, but cannot justify funding for such big projects. My modest department at the Texas A\&M University-Commerce has a much better secondary teacher education program than CERN can even dream of.  

The future of big science might be on what some still consider as being small science. The ELI light source under construction in Romania is an example which might answer some of the most important questions in particle and nuclear physics. Light is a wonderful tool, both on-shell and off-shell. One should not simply give funds to those who do not have any promising idea. When the concept of an idea has a strong scientific basis, investment is not futile.  I think it is time for particle physicists to shift to more practical grounds and include more light in their research\footnote{Or just give me the money. I have a couple of good ideas on how to use it wisely.}. And let there be light.

\begin{acknowledgments}
The author is grateful to Adeola Adeluyi, Gerhard Baur, Ted Barnes,  Kai Hencken, Spencer Klein, Michael Murray and and Farouk Aksouh  for beneficial discussions.  This work was partially supported by the U.S. DOE grant
DE-FG02-08ER41533 and U.S. NSF  Award No. PHY- 1415656.
\end{acknowledgments}

\end{document}